\documentclass[letterpaper, 10 pt, conference]{ieeeconf}  

\IEEEoverridecommandlockouts                              

\usepackage{graphics} 
\usepackage{lipsum} 
\usepackage{multicol} 
\usepackage{graphicx} 
\usepackage{amsmath}
\usepackage{graphicx}
\usepackage{lastpage}
\usepackage{optidef}
\usepackage{fancyhdr}
\usepackage{float}
\usepackage{nomencl} 
\usepackage{framed} 
\usepackage{mathrsfs}
\usepackage{amssymb}
\usepackage[ruled,vlined,linesnumbered]{algorithm2e}

\usepackage{balance}

\usepackage{tikz}
\usetikzlibrary{shapes.geometric, arrows.meta}
\usetikzlibrary{decorations.markings}
\usepackage{varwidth}
\usepackage{placeins}

\usepackage{stackengine}
\newcommand\barbelow[1]{\stackunder[1.2pt]{$#1$}{\rule{.8ex}{.075ex}}}

\usepackage{subfigure}

\usepackage{silence}
\usepackage{dblfloatfix}

\usepackage{fancyhdr}
\usepackage{lastpage}
\usepackage{mwe}
\usepackage{soul}
\soulregister\ref7
\soulregister\eqref7
\soulregister\cite7

\usepackage[finalnew]{trackchanges}
\addeditor{AN}
\addeditor{NJ}
\addeditor{HP}

\title{\LARGE \bf
Robust Control Co-Design with Receding-Horizon MPC}

\author{Austin L. Nash$^{1}$, Herschel C. Pangborn$^{2}$, and Neera Jain$^{3}$%
\thanks{*This work is supported by the U.S. Office of Naval Research Thermal Science and Engineering Program under contract number N000141712333.}
\thanks{$^{1}$Austin L. Nash is an Assistant Professor in the Mechanical Engineering Department,
        Kettering University, Flint, MI, USA. Email:
         {\tt\small anash@kettering.edu}}%
\thanks{$^{2}$Herschel C. Pangborn is an Assistant Professor in the Department of Mechanical Engineering,
       Pennsylvania State University, University Park, PA, USA. Email:
        {\tt\small hcpangborn@psu.edu}}%
 \thanks{$^{3}$Neera Jain is an Assistant Professor in the School of Mechanical Engineering,
       Purdue University, West Lafayette, IN, USA. Email:
        {\tt\small neerajain@purdue.edu}}%
}

\begin{document}

\raggedbottom

\maketitle
\thispagestyle{empty}
\pagestyle{empty}


\begin{abstract}

Control co-design (CCD) {is a technique for} improving the closed-loop performance of systems {through the coordinated design of} both plant parameters and an optimal control policy. While model predictive control (MPC) is an attractive {control} strategy for many systems, embedding {it} within a CCD algorithm presents challenges because obtaining a closed-form solution for this receding-horizon optimization strategy is often not feasible. This paper meets that challenge by including a robust MPC formulation within the inner loop of a CCD algorithm. As exemplified by application to an aircraft thermal management system, the proposed algorithm closely matches the plant design of an open-loop benchmark. However, unlike the open-loop approach, the proposed algorithm can leverage MPC control variables designed a priori to achieve robust online operation under disturbance profiles that differ from those used for design.

\end{abstract}


\section{Introduction} \label{sec-Introduction}

As performance requirements for a wide range of systems become more stringent, control co-design (CCD), also known as combined plant and control design, has been gaining renewed attention. This is due to the performance limitations of designing a controller only after the plant is designed, and thus the open loop dynamics are fixed. Instead, by considering both the design of plant parameters and an optimal control policy at the same time, more degrees of freedom are available to the engineer to achieve desired performance objectives. CCD has largely been limited to the design of static optimal control policies. However, this precludes the use of CCD with attractive control methods, such as model predictive control (MPC), which explicitly considers state and input constraints and uses a finite prediction horizon to take anticipatory control actions. Because MPC is often implemented as an online optimal control problem without a closed-form solution, there are significant challenges to embedding this control method within a CCD approach.

CCD is an interdisciplinary research area with contributions from engineering design and control experts alike.  Control systems researchers have emphasized the design of CCD methods that synthesize static optimal controllers, ranging from LQR \cite{jiang_iterative_2016,fathy_nested_2003,milman_combined_1991} to H-infinity \cite{nash_combined_2019,nash_hierarchical_2020}. In the design community, a major emphasis is on static parameter optimization and an open-loop control policy that assumes the future is known perfectly \cite{herber_problem_2019,allison_co-design_2014}. In \cite{diangelakis_process_2017}, Diangelakis et al. present a CCD algorithm with multi-parametric MPC that provides an explicit relationship between the control actions and design variables. However, explicit MPC approaches are typically limited to low-order systems and short horizons due to the computational complexity and memory storage requirements associated with a priori solutions of the control laws. To account for disturbance and process uncertainties, \cite{gutierrez_mpc-based_2014,sanchez-sanchez_simultaneous_2013,ricardez-sandoval_methodology_2011} use system identification techniques at each optimization step to quantify closed-loop variability. Ref. \cite{docimo_plant_2021} uses MPC within a simultaneous plant and controller optimization for a hybrid electric vehicle.

In this paper we present a control co-design algorithm for optimizing both plant parameters and a robust model predictive controller for systems with bounded additive uncertainty. The proposed algorithm incorporates a recursive MPC optimization within an inner loop without requiring a closed-form solution. This results in a set of optimal control variables and feedback gains that can be implemented in closed-loop \emph{without requiring online optimization}. We demonstrate the efficacy of the algorithm through a case study on a notional aircraft thermal management system. 

The remainder of the paper is organized as follows. In Sec. \ref{sec-Preliminaries} we derive the model of a {notional }thermal management system used to exemplify the proposed CCD algorithm. Sec. \ref{sec-CCD_rMPC} describes that algorithm{ which incorporates the design of a robust MPC controller.} In Sec. \ref{sec-OLCCD}, we present an open-loop CCD algorithm used as a baseline for comparison. Sec. \ref{sec-CaseStudies} illustrates the efficacy of the proposed algorithm in application to the thermal management system, and Sec. \ref{sec-Conclusions} concludes the paper.

\section{Preliminaries: Dual-Tank Thermal Management System} \label{sec-Preliminaries}

\begin{figure}[!htb]
\begin{center}
\includegraphics[width=0.75\linewidth]{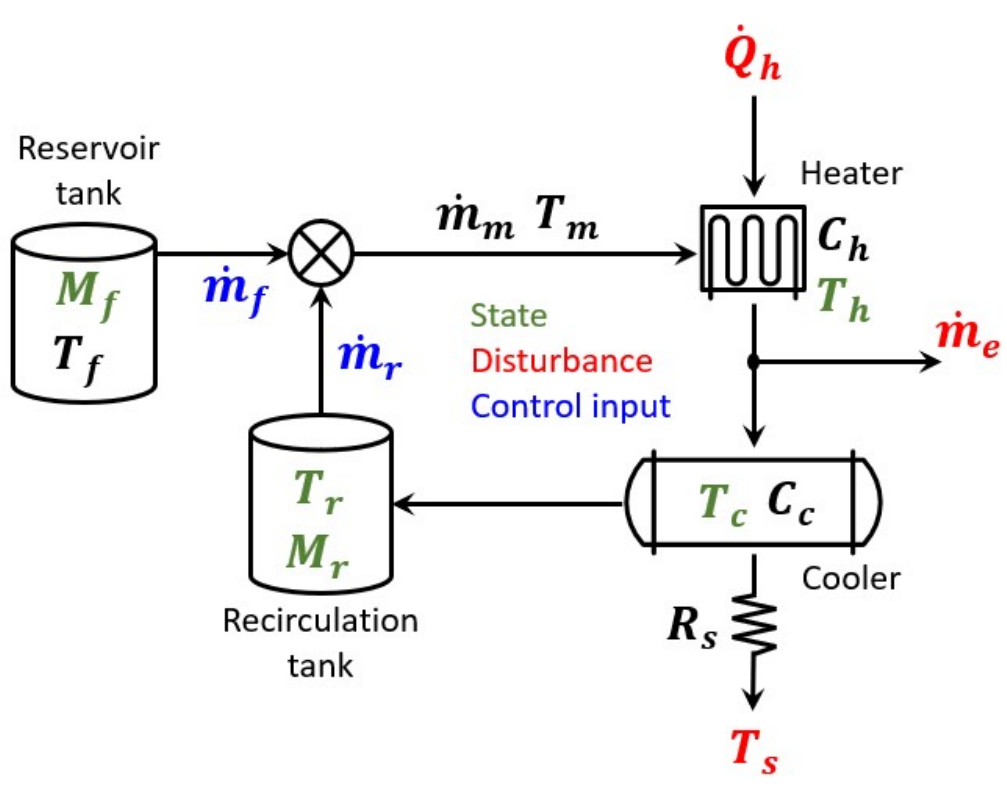}
\end{center}
\vspace*{-\baselineskip}
\caption{Schematic of a notional dual-tank aircraft thermal management system.}
\label{fig-two_tank_diagram}
\end{figure}

Here we derive the plant model that will be used throughout the rest of the paper. The dual-tank fuel thermal management system (plant) considered in this work and developed originally in \cite{doman_fuel_2018,huang_thermal_2018} is shown in Fig. \ref{fig-two_tank_diagram}. 

In the dual tank architecture, working fluid with mass $M_r$ and temperature $T_r$ exits a dedicated recirculation tank with controlled mass flow rate $\dot{m}_r$. Concurrently, working fluid with mass $M_f$ and temperature $T_f$ exits a secondary reservoir tank with controlled mass flow rate $\dot{m}_f$. The two controlled flows are mixed with a valve to produce downstream mass flow rate $\dot{m}_m$ and temperature $T_m$. The mixed working fluid absorbs energy from a pulsed heat load $\dot{Q}_h$ applied to a heater component with lumped working fluid temperature $T_h$ and thermal capacitance $C_h$. Downstream of the heater, working fluid is allowed to exit the cycle at mass flow rate $\dot{m}_e$; this mimics some aircraft thermal management systems wherein the working fluid is actually fuel for the engine and exits the system as needed for propulsion and power generation. The remaining working fluid flows into a cooler with lumped working fluid temperature $T_c$ and thermal capacitance $C_c$ where thermal energy is rejected to a secondary system at a temperature $T_s$ via resistance element $R_s$. The working fluid is then routed back into the recirculation tank to complete the thermal loop. 

Five dynamic states are sufficient to describe the first law dynamics of the system:  $M_f, \enspace M_r, \enspace T_h, \enspace T_c, \enspace \text{and} \enspace T_r$. The heat load $\dot{Q}_h$, secondary source temperature $T_s$, and exiting flow rate $\dot{m}_e$ are treated as exogenous disturbances while the mass flow rates from each tank, $\dot{m}_f$ and $\dot{m}_r$, are treated as control inputs. The state equations for the dual-tank system are given in Eq.~\eqref{eqn-state_eqs},

{\small
\begin{subequations}
\begin{align}
    \frac{dM_f}{dt} &= -\dot{m}_f \\
    \frac{dM_r}{dt} &= \dot{m}_f - \dot{m}_e \\
    C_h\frac{dT_h}{dt} &= \dot{m}_mc_p\left(T_m-T_h\right) + \dot{Q}_h \\
    C_c\frac{dT_c}{dt} &= \left(\dot{m}_m-\dot{m}_e\right)c_p\left(T_h-T_c\right) + \frac{1}{R_s}\left(T_s-T_c\right) \\
    M_r\frac{dT_r}{dt} &= \left(\dot{m}_m-\dot{m}_e\right)\left(T_c-T_r\right) ,
\end{align} 
\label{eqn-state_eqs}
\end{subequations}}

\noindent where $c_p$ is the specific heat capacity of the working fluid and the algebraic mixing equations are given by $\dot{m}_m = \dot{m}_f + \dot{m}_r$ and $\dot{m}_mT_m = \dot{m}_fT_f + \dot{m}_rT_r$.

We note that the parameters $C_h$, $C_c$, $T_f$, and $R_s$ are plant design variables that can be optimized along with a control policy in a CCD algorithm. The selection of these parameters captures tradeoffs between the mass/size and capability of individual components in the system. For example, a large value of $C_c$ corresponds to a large cooler component which may be advantageous from a thermal management perspective but disadvantageous from a system mass perspective. We also note that the dynamic state describing the mass of fluid in the reservoir tank, $M_f$, is an uncontrollable state in the dual-tank topology. Therefore, the robust MPC algorithm formulated in this paper must leverage the controllable subsystem of the plant.

In the following section, we derive our robust control co-design (rCCD) algorithm that \emph{directly leverages a receding-horizon robust MPC formulation in the CCD algorithm itself}, thereby enabling the design of an optimal plant/controller design that is more robust to uncertainties in the expected load profile than designs resulting from conventional open-loop algorithms in the literature.

\section{Robust Control Co-Design with Receding-horizon MPC} \label{sec-CCD_rMPC}

Here we describe our proposed robust control co-design (rCCD) algorithm, which includes a robust MPC (rMPC) formulation nested within the algorithm. We note that while the formulation here is aimed at optimizing the plant design and control policy of a dual-tank fuel thermal management system for minimum fluid mass, the proposed formulation is widely applicable to a range of actively-controlled dynamic systems.

\subsection{Algorithm Description}
Our rCCD algorithm directly incorporates a receding-horizon rMPC synthesis into a CCD algorithm. In other words, we solve a receding-horizon rMPC problem at each time step of a desired performance profile. We discretize the entire time horizon into $n_t = t_f/\tau _s$ time steps where $t_f$ is the final time and $\tau _s$ is the control sample rate. We treat the initial state variables of the system, $x_0$, and the initial control variables of the system, $u_0$, as decision variables. Plant design parameters $C_h$, $C_c$, $T_f$, and $R_s$ are also treated as decision variables in vector $p$. We define the optimization problem based on a nominal disturbance profile $\mathbb{D} = [d_{k=1}, \enspace d_2, \enspace \hdots, \enspace d_{n_t}]$, where $d_k \in \mathbb{R}^{n_d} $ is the vector of disturbances at the $k^{th}$ time step and $n_d$ is the number of exogenous disturbances acting on the system. In addition to this nominal disturbance, we also assume there exists a bounded uncertainty in the disturbances given by the set $\mathcal{W}:= \enspace \left\{ w \in \mathbb{R}^{n_d} | \enspace \barbelow{w} \leq w \leq \bar{w} \right\}$. Hence the true disturbance at the $k^{th}$ time step is given by $\tilde{d}_k = d_k + w_k$. The symbols $\bar{\alpha}$ and $\barbelow{\alpha}$ represent upper and lower bounds on the variable $\alpha$.

\begin{figure}[!htb]
\begin{center}
\includegraphics[width=0.91\linewidth]{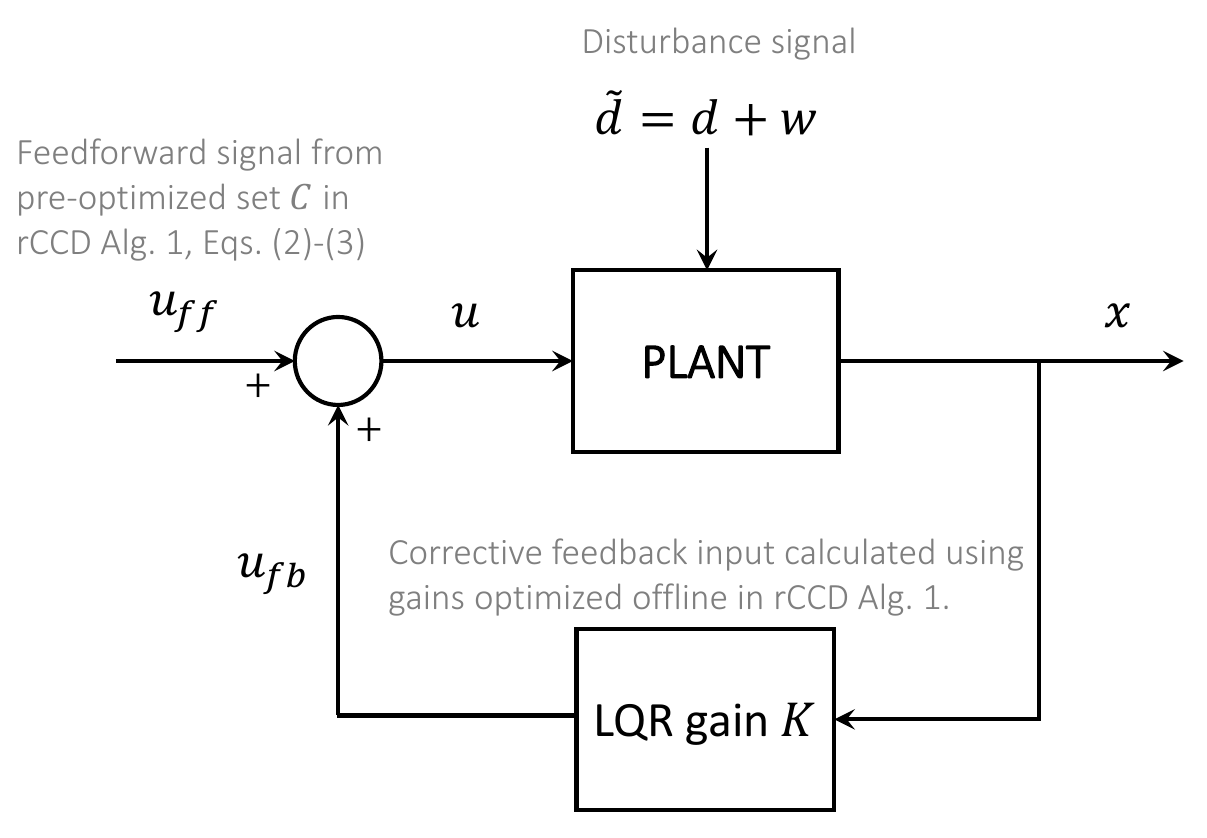}
\end{center}
\vspace*{-\baselineskip} 
\caption{Block diagram illustration of the plant and control elements optimized in the proposed rCCD algorithm.}
\vspace*{-\baselineskip}
\label{fig-M3}
\end{figure}

The objective of our rCCD algorithm is to minimize the initial mass of working fluid in the dual-tank system. We note that in applications such as aircraft thermal management, lower mass generally correlates with desired performance criteria such as increased fuel efficiency. The full statement of the rCCD optimization problem is shown in Eq.~\eqref{eqn-rCCD} and the algorithm is described by the pseudo code in Alg.~\ref{alg-rCCD}. The rMPC formulation nested within each iteration of the rCCD algorithm is stated in Eq.~\eqref{eqn-rMPC}. Additionally, Fig.~\ref{fig-M3} provides a block diagram representation of the interaction between the plant and controller design elements optimized in the rCCD algorithm.
\vspace*{-\baselineskip}

{\small
\begin{equation}
\begin{aligned}
\min_{x_0,u_0,p} \quad & J_{sys}  \\
\textrm{s.t.} \quad & h_k = 0 \enspace \forall \enspace k \in [1,n_t] \\
& x_{k+1} = f\left(x_k,u_k,d_k,t_k\right)  \in \mathcal{X} \enspace \forall \enspace k \in [1,n_t] \\
& \mathcal{X}:= \enspace \left\{ x \in \mathbb{R}^{5} | \enspace \barbelow{x} \leq x \leq \bar{x} \right\}
\end{aligned}
\label{eqn-rCCD}
\end{equation}}

\noindent where

{\small
\begin{equation}
\begin{aligned}
& J_{sys} = M_{sys} = \frac{1}{c_p}\left({C_c + C_h}\right) + M_{f,0} + M_{r,0} \\
& p = [C_c, \enspace C_h, \enspace T_f, \enspace R_s]^T \\
& x_0 = x_{t=0} = [M_{f,0}, \enspace M_{r,0}, \enspace T_{h,0}, \enspace T_{c,0}, \enspace T_{c,0}]^T \\
& u_0 = u_{t=0} = [\dot{m}_{f,0}, \enspace \dot{m}_{r,0}]^T \\
& h_k:= \text{rMPC (Eq.~\eqref{eqn-rMPC}) feasibility maps to } \{0,1\} \\
& u_k:= \text{synthesized in nested rMPC formulation in Eq.~\eqref{eqn-rMPC}}
\end{aligned} \nonumber 
\end{equation}}
\smallskip

\normalsize
We initialize the algorithm by setting the outer loop iteration index to $j=1$ and assigning initial guesses for the plant variables $p^{(j)}$, initial state variables $x_0^{(j)}$, and initial control variables $u_0^{(j)}$. We then evaluate the objective function value $J_{sys}$ for the $j^{th}$ iteration. Recall that for our algorithm, $J_{sys} = M_{sys}$ is the amount of working fluid in the system at time $t=0$.

We then enter the inner loop of the rCCD algorithm, which uses the receding-horizon rMPC formulation in Eq.~\eqref{eqn-rMPC} to evaluate robust feasibility for the plant/controller design combination in outer loop iteration $j$. The nested rMPC problem is a robust min-max MPC formulation \cite{lofberg_approximations_2003} that attempts to ensure constraint satisfaction for disturbances $\tilde{d}_i = d_i + w_i$ with $w_i \in \mathcal{W}$ for the $N_p$-length receding horizon. The decision variables within the rMPC algorithm are open-loop control variables $C = [c_{i=1}, \enspace c_2, \enspace \hdots, \enspace c_{N_p}]$ and the disturbance uncertainty variables $W = [w_{i=1}, \enspace w_2,\enspace \hdots, \enspace w_{N_p}]$. The objective of the rMPC formulation is to minimize the total control effort at each time step subject to constraints that ensure the state variables $x_i$ and control variables $u_i$ remain within the sets $\mathcal{X}$ and $\mathcal{U}$, respectively, for all uncertain disturbance variables $w_i$ for which $w_i \in \mathcal{W}$ holds. The state $M_f$, which is uncontrollable, is also constrained to remain within a bounded region. We note that the rMPC algorithm has full preview of the planned disturbances for the $N_p$ steps in the prediction horizon.

The inner loop is initialized by setting the inner loop iteration index to $k=1$ and stepping through each time step $k=[1,n_t]$. The process begins by assigning the (expected) disturbance vector $d_i$ for each $i^{th}$ time sample in the $N_p$-length prediction horizon. We then linearize about the current state, discretize the linear dynamics, and extract the controllable subsystem. We note that the linearization at each time step is performed to maintain an accurate control model since the masses in the two tanks change considerably during operation. We use the resulting discrete-time controllable system to solve for a static LQR gain matrix $K_k$ for the $k^{th}$ time step. The gain matrix is then passed to the rMPC problem. If the rMPC problem has a feasible solution for the $k^{th}$ time step, we assign the outer loop constraint variable for that time step, $h_k$, a zero value. Alternatively, if the rMPC problem has an infeasible solution for the $k^{th}$ time step, $h_k$ is assigned a value of one. In future work we will seek to incorporate guarantees of closed-loop stability and recursive feasibility under this successive linearization \cite{henson_nonlinear_1998}.

For a feasible time step $k$, we store the open-loop control variables for the first step of the rMPC horizon, $c_{i=1}$, and the static gains $K_k$, in the sets $\mathbb{C}$ and $\mathbb{K}$, respectively. The total control input $u_k$ for the time step is computed based on the optimal open-loop control variables from the rMPC problem, the static gains for the time step, and the linearized point (denoted by superscript $e$ in Alg.~\ref{alg-rCCD}). We then simulate the continuous-time, nonlinear plant (see Eq.~\eqref{eqn-state_eqs}) forward from $t_k$ to $t_{k+1}$ using the $k^{th}$ state $x_k$, control input $u_k$, and expected disturbance $d_{k}$ to obtain the state $x_{k+1}$ at time $t_{k+1}$. Outer loop iteration $j$ is deemed feasible if and only if $h_k = 0 \enspace \text{and} \enspace x_{k+1} \in \mathcal{X} \enspace \forall \enspace k \in [1,n_t]$; in other words, we require both a feasible rMPC problem and a subsequent state $x_{k+1}$ (simulated on the nonlinear plant) within the allowable set $\mathcal{X}$ for each time step $k$. Once feasibility has been evaluated, the decision variables are updated for iteration $j+1$ and the process is repeated until the tolerance $\varepsilon$ for the user-chosen optimization solver is satisfied.

\begin{algorithm}[!htb]
{\small\SetAlgoLined
 initialize decision variables for outer loop with iteration index $j$\;
 $j\gets 1$, $p^{(j)} \gets p^{(1)}$, $x_0^{(j)} \gets x_0^{(1)}$, $u_0^{(j)} \gets u_0^{(1)}$\;
 initialize tolerance variable, $\Delta J \gets 1$\;
 \While {$\Delta J \geq \varepsilon$ \text{(iterate until convergence)}}{
 compute $J_{sys}^{(j)}$ for current iteration $j$ using $p^{(j)}$\;
 $k \gets 1$\;
 $x_k \gets x_0^{(j)}$, $u_k \gets u_0^{(j)}$, $\mathbb{C}^{(j)} \gets \emptyset$, $\mathbb{K}^{(j)} \gets \emptyset$\;
 \For {$k = 1:n_t$}{
  assign disturbance signal $d_i \enspace \forall \enspace i\in \left[k,k+N_p\right]$\;
  linearize about current point $\left( x,u,d\right)$\;
  discretize dynamics, extract controllable system\;
  compute LQR gain matrix for $k^{th}$ step, $K_k$\;
  evaluate feasibility of rMPC problem in Eq.~\eqref{eqn-rMPC}\;
  \eIf{Eq.~\eqref{eqn-rMPC} has feasible solution}{
   $h_k \gets 0$, (feasible time step $k$)\;
   $\mathbb{C}^{(j)} \gets \left\{\mathbb{C}^{(j)}, \enspace c_{i=1}\right\}$, $c_{i=1}$ from Eq.~\eqref{eqn-rMPC}\;
   $\mathbb{K}^{(j)} \gets \left\{\mathbb{K}^{(j)}, \enspace K_k\right\}$\;    
   $u_k \gets c_{i=1} + K_k \left(x_k - x^e\right) + u^e$\;
  simulate nonlinear plant over $[t_k,t_{k+1}]$, to get $x_{k+1}$\;     
   }{
   $h_k \gets 1$, (infeasible time step)\;
   if here, outer loop iteration $j$ is infeasible\;
  }
  $k\gets k+1$\;
 }
 \eIf{$h_k = 0 \enspace \text{\&} \enspace x_{k+1} \in \mathcal{X} \enspace \forall \enspace k \in [1,n_t]$ }{
 iteration $j$ feasible, solver will compute current tolerance $\Delta J$ and update decision variables to converge toward smaller objective\;
  $j\gets j + 1$\;
 update decision variables $p^{(j)}$, $x_0^{(j)}$, $u_0^{(j)}$\;
 }{
 iteration $j$ infeasible, solver will update decision variables as needed to converge toward feasible solution\;
  $j\gets j + 1$\;
 update decision variables $p^{(j)}$, $x_0^{(j)}$, $u_0^{(j)}$\;
 }
 }}
 \caption{Proposed rCCD algorithm}
 \label{alg-rCCD}
\end{algorithm}

{\small
\begin{equation}
\begin{aligned}
\min_{C} \max_{W} \quad & J_{u} = \sum_{i=1}^{N_p} c_i^TRc_i \\
\textrm{s.t.} \quad & x_{i+1}  = Ax_i + B_2u_i + B_1\tilde{d} \in \mathcal{X} \enspace \forall \enspace i \in [1,N_p] \\
& \barbelow{M}_f \leq M_{f,i+1} \leq \bar{M}_f \enspace \forall \enspace i \in [1,N_p] \\
& u_{i} \in \mathcal{U}  \enspace \forall \enspace i \in [1,N_p] \\
& w_{i} \in \mathcal{W}  \enspace \forall \enspace i \in [1,N_p] \\
& \mathcal{X}:= \enspace \left\{ x \in \mathbb{R}^{4} | \enspace \barbelow{x} \leq x \leq \bar{x} \right\} \\
& \mathcal{U}:= \enspace \left\{ u \in \mathbb{R}^{2} | \enspace \barbelow{u} \leq u  \leq \bar{u}, \enspace |u_i - u_{i-1}| \leq \delta \bar{u} \right\} \\
& \mathcal{W}:= \enspace \left\{ w \in \mathbb{R}^{3} | \enspace \barbelow{w} \leq w \leq \bar{w} \right\} 
\end{aligned}
\label{eqn-rMPC}
\end{equation}}
\smallskip

\noindent where

{\small
\begin{equation}
\begin{aligned}
& C = [c_{i=1}, \enspace c_2,\enspace \hdots, \enspace c_{N_p}] \\
& W = [w_{i=1}, \enspace w_2,\enspace \hdots, \enspace w_{N_p}] \\
& u_i = c_i + K_k\left(x_i-x^e\right) + u^e \\
& x = [M_r, \enspace T_h, \enspace T_c ,\enspace T_r]^T \\
& u = [\dot{m}_f, \enspace \dot{m}_r]^T \\
& d = [\dot{Q}_h, \enspace T_s, \enspace \dot{m}_e]^T \\
& \tilde{d}_i = d_i + w_i \\
& M_{f,i+1} = M_{f,i} - \dot{m}_{f,i}\cdot \tau _s 
\end{aligned} \nonumber 
\end{equation}}
\smallskip

\noindent The optimal solution to the rCCD algorithm then outputs the following variables and signals:

\bigskip
\begin{itemize}
    \item optimal vector of plant variables $p^*$
    \item optimal initial state $x_0^*$
    \item optimal initial control variables $u_0^*$
    \item optimal sample-scheduled set of open-loop control variables stored in $\mathbb{C}^* = \{c_{k=1}, \enspace c_2, \enspace \hdots, \enspace c_{n_t}\}$
    \item optimal sample-scheduled set of static LQR gains stored in $\mathbb{K}^* = \{K_{k=1}, \enspace K_2, \enspace \hdots, \enspace K_{n_t}\}$.
\end{itemize}
\bigskip

The resulting optimal system leverages both feedforward and feedback control action from the inner loop receding-horizon rMPC formulation. While theoretical guarantees of recursive feasibility under disturbance uncertainty are left to future work, the use of methods from the robust MPC literature allows the CCD algorithm to design a controller that inherently seeks to compensate for disturbance uncertainty, $\tilde{d} = d + w$ such that $w\in\mathcal{W}$, in maintaining constraints.
In the next section, we describe a conventional CCD algorithm that will be used as a benchmark to evaluate the efficacy of our proposed rCCD algorithm.

\section{Baseline for Comparison: Open-loop Control Co-design} \label{sec-OLCCD}

\begin{figure}[!htb]
\begin{center}
\includegraphics[width=0.65\linewidth]{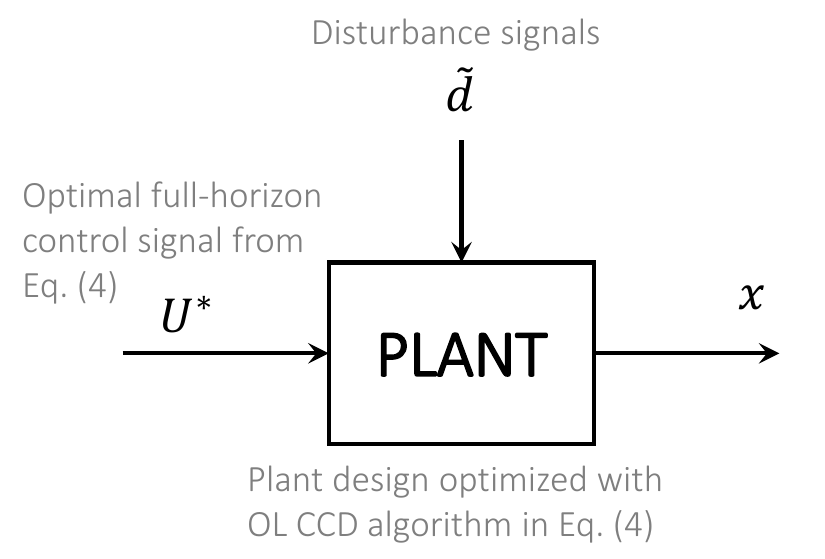}
\end{center}
\vspace*{-\baselineskip}
\caption{Block diagram illustration of plant/control elements optimized in a conventional OL CCD algorithm.}
\label{fig-M1}
\end{figure}

\noindent Direct transcription, or DT, is a popular CCD technique in the literature wherein a system's trajectory is discretized into time steps, and the state and control variables at each time step are treated as decision variables in a CCD algorithm \cite{herber_problem_2019,allison_special_2014,peddada_optimal_2019}. The DT CCD problem can then be solved as a nonlinear program where the system state equations are imposed as equality constraints. Here, we will use a similar open-loop, all-at-once style CCD algorithm as a baseline against which to benchmark our proposed rCCD algorithm. We again discretize the entire time horizon into $n_t = t_f/\tau _s$ time steps where $t_f$ is the final time and $\tau _s$ is the control sample rate. We treat the vector of initial state variables of the system $x_0$ and the matrix $\mathbb{U}$ of control variables at each $k^{th}$ time step $u_k$, denoted as $\mathbb{U} = [u_{k=1}, \enspace u_2, \enspace \hdots, \enspace u_{n_t}]$, as decision variables. The plant parameters $C_h$, $C_c$, $T_f$, and $R_s$ are again treated as decision variables in vector $p$. The full statement of the open-loop CCD optimization problem is given in Eq.~\eqref{eqn-OL_CCD}; Fig.~\ref{fig-M1} provides a block diagram representation of the plant and control elements designed using the baseline OL CCD algorithm.

{\small
\begin{equation}
\begin{aligned}
\min_{x_0,\mathbb{U},p} \quad & J_{sys} \\
\textrm{s.t.} \quad & x_{k+1} \in \mathcal{X} \enspace \forall \enspace k \in [1,n_t] \\
& u_{k} \in \mathcal{U}  \enspace \forall \enspace k \in [1,n_t] \
& \mathcal{X}:= \enspace \left\{ x \in \mathbb{R}^{5} | \enspace \barbelow{x} \leq x \leq \bar{x} \right\} \\
& \mathcal{U}:= \enspace \left\{ u \in \mathbb{R}^{2} | \enspace \barbelow{u} \leq u  \leq \bar{u}, \enspace |u_k - u_{k-1}| \leq \delta \bar{u} \right\} \\
\end{aligned}
\label{eqn-OL_CCD}
\end{equation}}
\smallskip

\noindent where

{\small
\begin{equation}
\begin{aligned}
& J_{sys} = M_{sys} = \frac{1}{c_p}\left({C_c + C_h}\right) + M_{f,0} + M_{r,0} \\
& \mathbf{\mathbb{U}} = [u_{k=1}, \enspace u_2, \hdots, \enspace u_{n_t}] \\
& p = [C_c, \enspace C_h, \enspace T_f, \enspace R_s]^T \\
& x = [M_f, \enspace M_r, \enspace T_h, \enspace T_c ,\enspace T_r]^T \\
& u = [\dot{m}_f, \enspace \dot{m}_r]^T \\
& d = [\dot{Q}_h, \enspace T_s, \enspace \dot{m}_e]^T \\
& x_0 = x_{t=0} = [M_{f,0}, \enspace M_{r,0}, \enspace T_{h,0}, \enspace T_{c,0}, \enspace T_{c,0}]^T \\
& x_{k+1} \enspace \text{from simulating nonlinear plant over $[t_k, t_k+\tau _s]$} 
\end{aligned} \nonumber
\end{equation}}
\smallskip

Similar to the rCCD algorithm, the objective of the OL CCD algorithm in Eq.~\eqref{eqn-OL_CCD}, $J_{sys}$, is to minimize the initial mass of working fluid in the dual-tank system, $M_{sys}$. At the $k^{th}$ discretized time step, the control variables $u_k$ are decision variables and the state variables $x_k$ are functions of the initial state $x_0$ and control and disturbance variables $u_i$ and $d_i$ for all past time steps $i \enspace \in \enspace [1,k]$. We impose constraints for each time step to ensure that the control variables $u_k$ are within the bounded set $\mathcal{U}$ and the state variables $x_{k+1}$ are within the bounded set $\mathcal{X}$, where $\mathcal{U}$ and $\mathcal{X}$ are defined in Eq.~\eqref{eqn-OL_CCD} and the quantity $\delta \bar{u}$ represents a maximum allowable change in the control variables. Subsequent state variables $x_{k+1}$ are again computed by simulating the continuous-time nonlinear plant, which is a function of $\left(x_k,u_k,d_k,t_k\right)$, forward in time from time $t_k$ to $t_k+\tau_s $. As with conventional CCD algorithms in the literature, we assume the algorithm has full access, or full preview, to a disturbance profile $\mathbb{D} = [d_{k=1}, \enspace d_2, \enspace \hdots, \enspace d_{n_t}]$. The outputs of the OL CCD algorithm are (1) an optimal set of plant parameters $p^*$, (2) an optimal initial state $x_0^*$, and (3) the optimal set of control variables for the entire performance profile $\mathbb{U} = [u_{k=1}, \enspace u_2, \enspace \hdots, \enspace u_{n_t}]$.

A main drawback of this class of DT-style open-loop CCD algorithms is they are susceptible to uncertainty in the exogenous disturbance signals; in other words, the optimal solution is reliant on an assumed disturbance profile. However, a fully optimized trajectory for one load (disturbance) profile may be ill-equipped to handle other load profiles. In the following section, we use a set of case studies to illustrate how our proposed rCCD algorithm can meet constraints in the presence of uncertainty in the load profile. 

\section{Results:  CCD Case Studies} \label{sec-CaseStudies}

In this section, we present a set of case studies to illustrate the efficacy of the proposed rCCD algorithm.  We use the conventional OL CCD strategy presented in Sec.~\ref{sec-OLCCD} as a benchmark against which to compare. 

\subsection{Optimal System Design Comparison}

For the case studies and results in this work, each algorithm is solved using the NOMAD nonlinear optimization algorithm, which interfaces with MATLAB through the open-source OPTI toolbox \cite{currie_opti_2012}. The optimizations were performed on a 32-core machine with a 3.0 GHz processor and 32 GB RAM. The rMPC problem nested in the rCCD algorithm is solved using the IPOPT solver interfaced with YALMIP \cite{lofberg_yalmip_2004}. We use water as the working fluid with a mission time (full time horizon) of 100 seconds and a one second control sample time for each algorithm. Table~\ref{tab-opt_vars} summarizes the optimal plant designs and optimal initial states, along with the enforced upper and lower bounds for each quantity, for both the OL CCD algorithm and the rCCD algorithm. We note that optimal control variables are not presented in Table~\ref{tab-opt_vars} for brevity; however, they will be presented graphically in subsequent subsections.

\begin{table}[!htb]
\caption{Comparison of optimal system designs and characteristics: OL CCD vs. proposed rCCD.}
\begin{center}
  \begin{tabular}{l l l l l l}
  \hline
  \noalign{\vskip 1mm} 
    \textbf{Design} & Units & Lower & Upper & \textbf{OL CCD} & \textbf{rCCD} \\
    \textbf{variable} & & bound & bound & &  \\
  \noalign{\vskip 1mm} 
  \hline
  \noalign{\vskip 1mm} 
    $\mathbf{C_c}$ & kJ/K & $5.0$ & $20.0$ & $\mathbf{5.0}$ & $\mathbf{5.0}$ \\ 
    $\mathbf{C_h}$    & kJ/K & $5.0$ & $20.0$ & $
    \mathbf{5.0}$ & $\mathbf{5.0}$ \\ 
    $\mathbf{R_s}$     & K/kW & $4.0$ & $5.0$ & $\mathbf{4.6}$ & $\mathbf{4.9}$ \\ 
    $\mathbf{T_f}$   & C & $20.0$ & $30.0$ & $\mathbf{20.0}$ & $\mathbf{20.0}$ \\ 
  \noalign{\vskip 1mm} 
  \hline
  \noalign{\vskip 1mm}     
    $\mathbf{M_{f,0}}$   & kg & $0.10$ & $10.0$ & $\mathbf{0.32}$ & $\mathbf{0.35}$  \\ 
    $\mathbf{M_{r,0}}$   & kg & $0.10$ & $10.0$ & $\mathbf{0.10}$ & $\mathbf{0.10}$ \\ 
    $\mathbf{T_{h,0}}$   & C & $45.0$ & $50.0$ & $\mathbf{49.9}$ & $\mathbf{48.0}$ \\ 
    $\mathbf{T_{c,0}}$   & C & $7.5$ & $50.0$ & $\mathbf{34.7}$ & $\mathbf{7.5}$  \\ 
    $\mathbf{T_{r,0}}$   & C & $7.5$ & $50.0$ & $\mathbf{19.6}$ & $\mathbf{7.5}$  \\ 
    \noalign{\vskip 1mm} 
    \hline
    \noalign{\vskip 1mm} 
    \multicolumn{4}{l}{\textbf{$J^* = M_{sys}^*$ (kg)}} & $\mathbf{2.81}$ & $\mathbf{2.84}$  \\ 
  \noalign{\vskip 1mm} 
  \hline
  \end{tabular}
\end{center}
\label{tab-opt_vars}
\end{table}

Table~\ref{tab-opt_vars} shows that the optimal objective function values are very similar for each algorithm. The minimum total working fluid mass that meets all constraints for the rCCD case is $M_{sys}^* = 2.84$ kg, which is just $1.0$\% higher than the optimal value $M_{sys}^* = 2.81$ kg for the OL CCD case. We note that, in either case, several of the optimal decision variables take on the lower bounds of the respective variables. The lower bounds of the plant decision variables were chosen such that the lumped parameter quantities here accurately represent the characteristics of high-fidelity physical components such as those optimized in \cite{nash_hierarchical_2020}. The benefit of the optimal system design resulting from our rCCD algorithm, however, is that the plant/controller combination is designed to be robust to uncertainties in the load profile for disturbances $\tilde{d} = d + w$ for all $w\in \mathcal{W}$. Conversely, the optimal design in the OL CCD is geared toward a specific load profile and is not optimized to be robust to load uncertainty. The case studies in the following subsections will demonstrate this through simulation.

\subsection{Case Study $1$: Perfect Disturbance Knowledge}

\begin{figure}
    \centering
    \subfigure[Heat load and exiting mass flow disturbance signals.]
    {
        \includegraphics[width=0.45\textwidth]{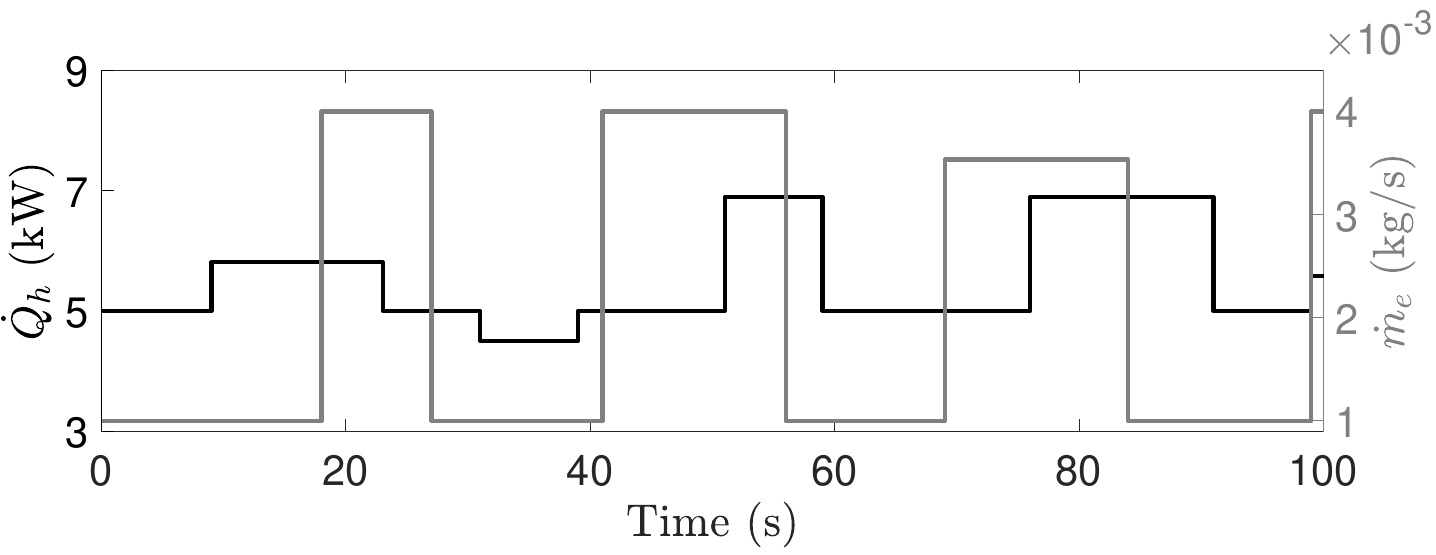}
        \label{fig-cs1_d}
    }
    \subfigure[Control input signal: reservoir tank mass flow rate.]
    {
        \includegraphics[width=0.45\textwidth]{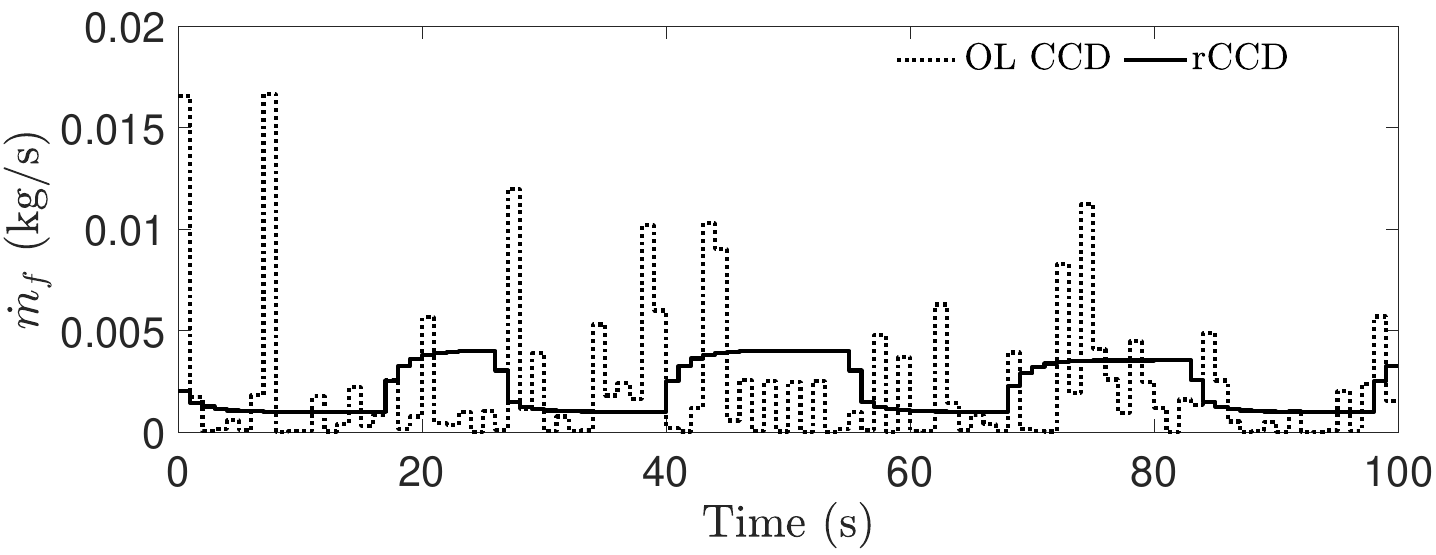}
        \label{fig-cs1_u1}
    }
    \subfigure[Control input signal: recirculation tank mass flow rate.]
    {
        \includegraphics[width=0.45\textwidth]{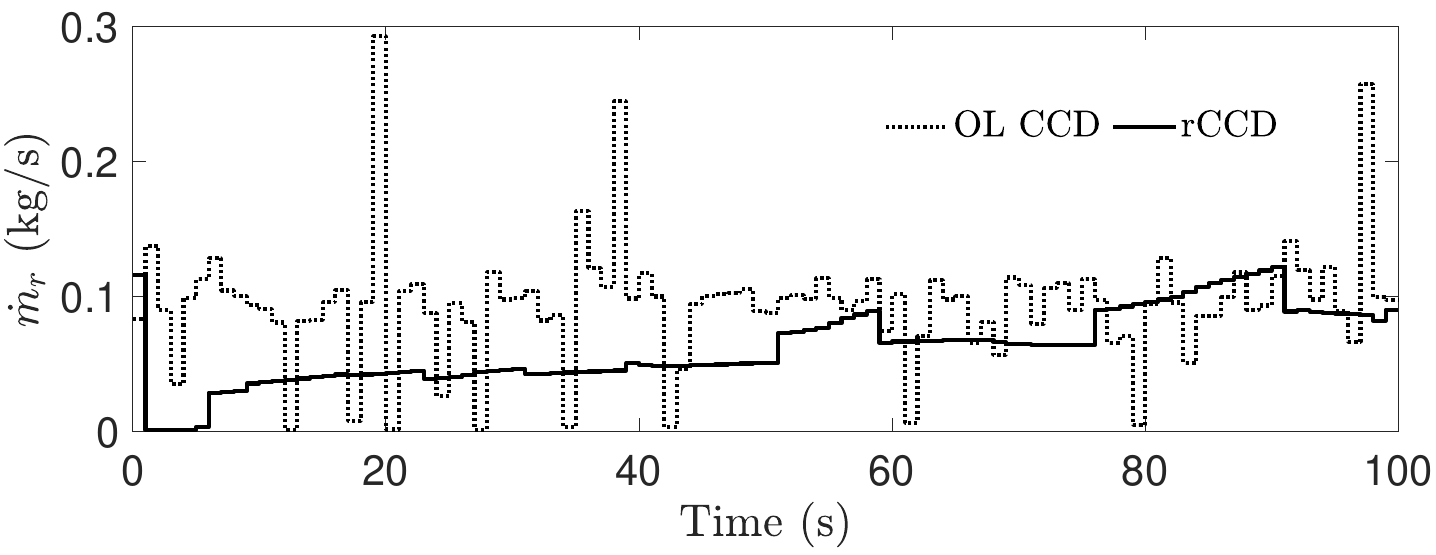}
        \label{fig-cs1_u2}
    }
    \subfigure[Heater temperature state (red dashed lines are constraint bounds).]
    {
        \includegraphics[width=0.45\textwidth]{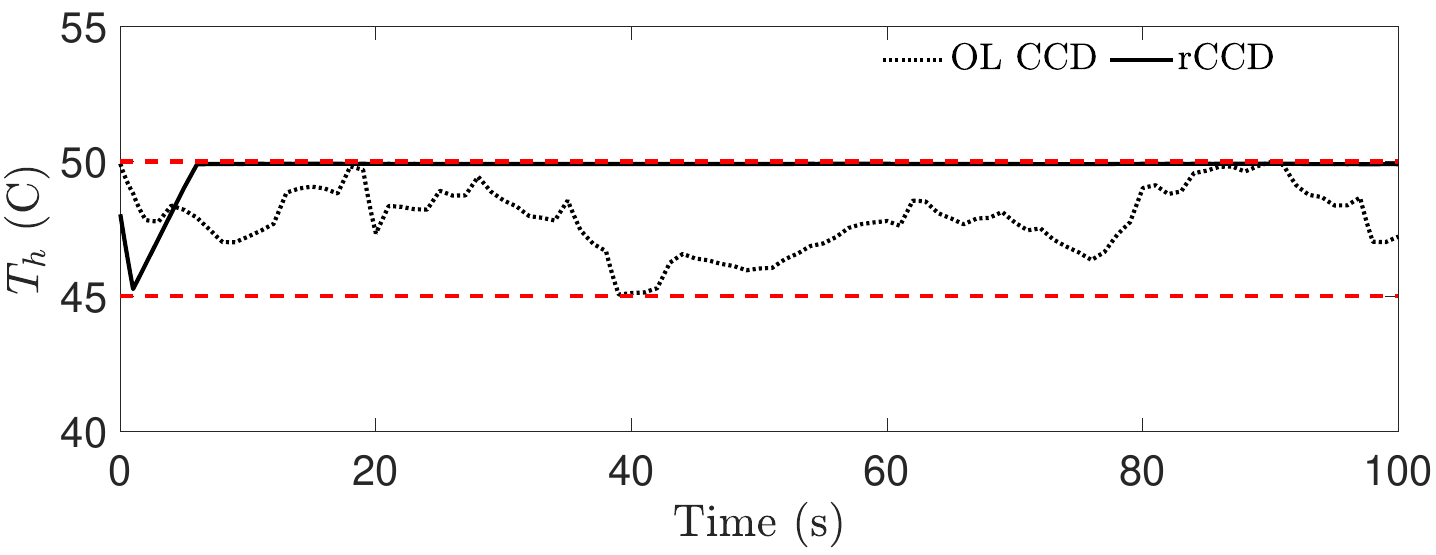}
        \label{fig-cs1_t}
    }
    \caption{\textbf{Case Study 1},  perfect disturbance knowledge:  both OL CCD and rCCD plant/controller designs meet the constraints with perfect knowledge of the load profile.}
    \label{fig-cs1}
\end{figure}

We first examine the optimal system responses assuming perfect disturbance knowledge at the system-build stage. In other words, the actual load profile is the exact profile the systems were optimized for. The simulated results are shown in Fig.~\ref{fig-cs1}.

Figure~\ref{fig-cs1_d} shows profiles for the heat load $\dot{Q}_h$ and mass flow rate exiting the thermal loop $\dot{m}_e$. Figures~\ref{fig-cs1_u1}-\ref{fig-cs1_u2} show the optimal control signals resulting from the rCCD and OL CCD algorithms, and Fig.~\ref{fig-cs1_t} shows the heater outlet temperature state $T_h$ for each algorithm. We see from Fig.~\ref{fig-cs1_t} that each algorithm satisfies the temperature constraints for the heater outlet temperature. In this case, the OL CCD algorithm represents the best possible initial fluid mass $M_{sys}$ that will permit a controller to satisfy the constraints due to its full knowledge of the load profile. Recall that the inner loop rMPC problem in the rCCD algorithm was formulated to minimize total control effort. The rCCD algorithm leverages both its feedforward and feedback control action to meet the constraints while conserving mass flow (control effort) from each tank. We note that even though each algorithm satisfies the heater outlet temperature constraints in Case Study 1, the optimal control signals are still different in the OL CCD and rCCD cases due to the fundamental differences between the OL CCD algorithm's open-loop control policy and the rCCD algorithm's recursive closed-loop (feedback) control policy.

\subsection{Case Study $2$: Unplanned Load Profile}

\begin{figure}
    \centering
    \subfigure[Heat load and exiting mass flow disturbance signals.]
    {
        \includegraphics[width=0.45\textwidth]{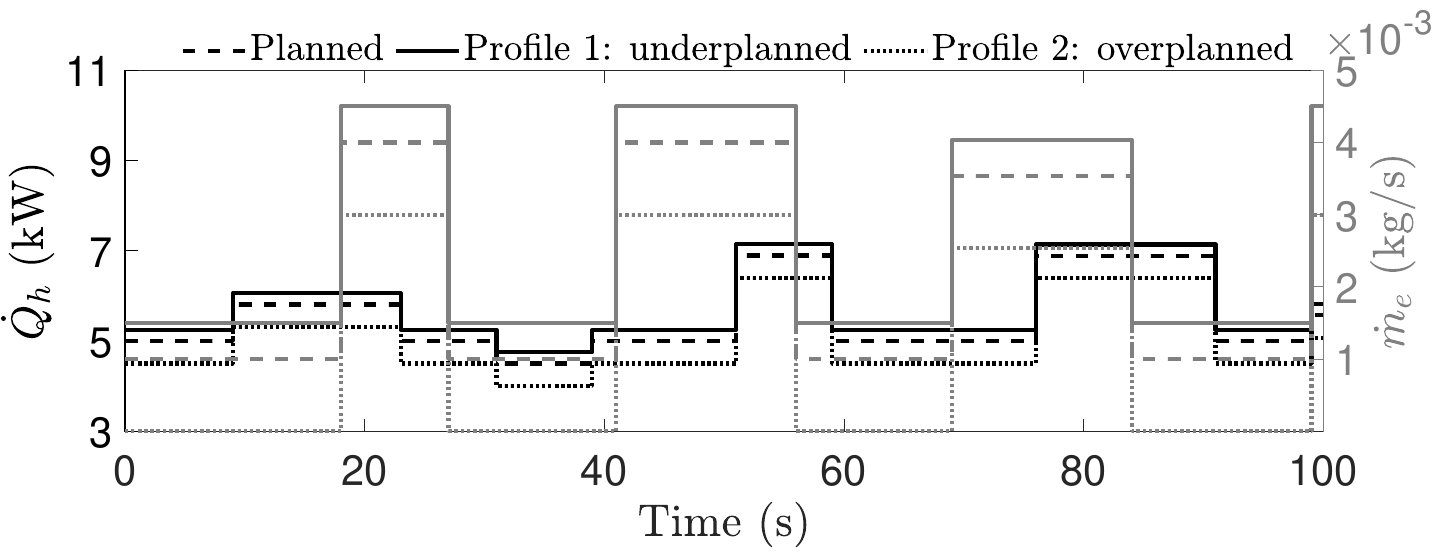}
        \label{fig-cs2_d}
    }
    \subfigure[Control input signal: reservoir tank mass flow rate.]
    {
        \includegraphics[width=0.45\textwidth]{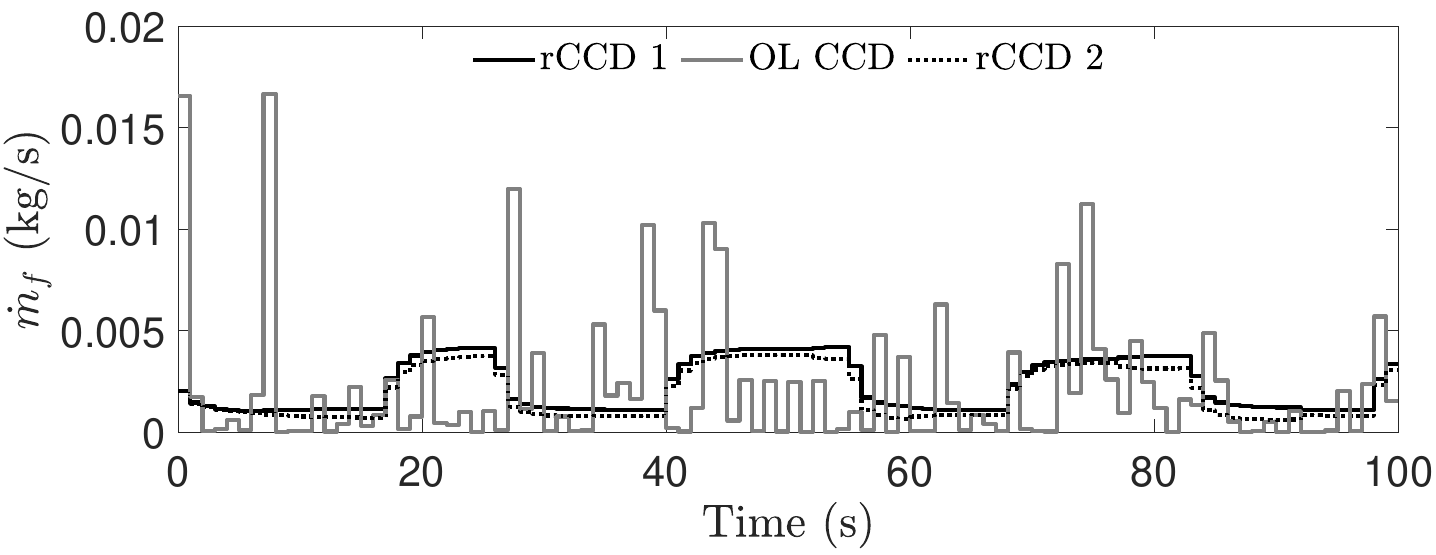}
        \label{fig-cs2_u1}
    }
    \subfigure[Control input signal: recirculation tank mass flow rate.]
    {
        \includegraphics[width=0.45\textwidth]{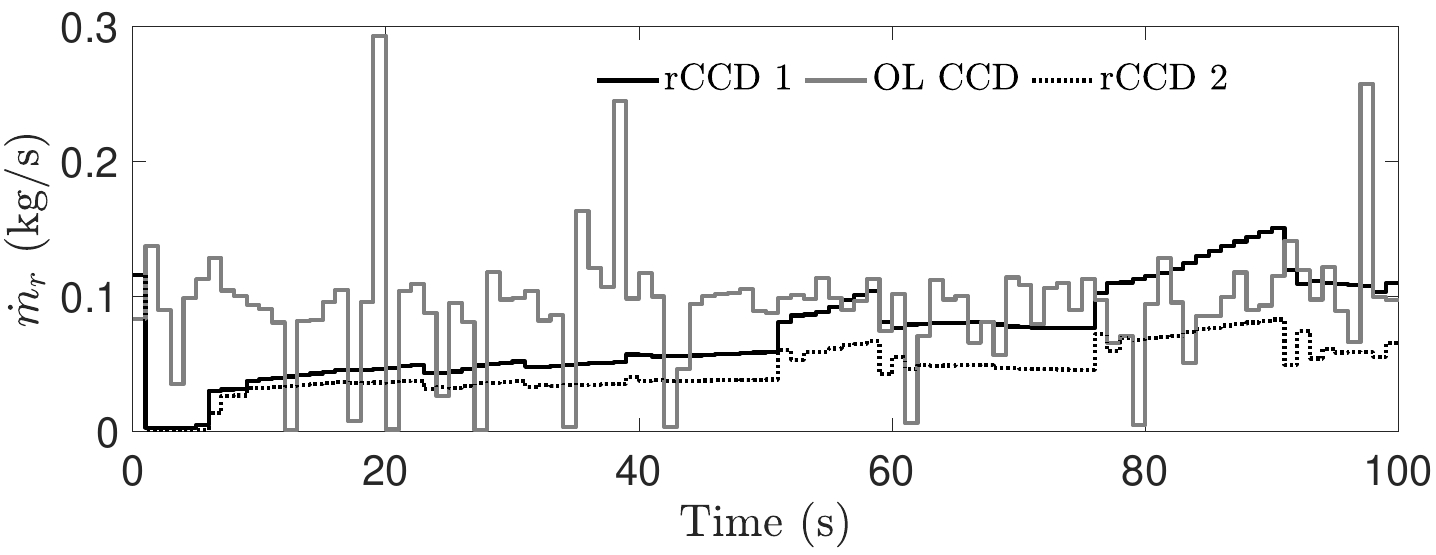}
        \label{fig-cs2_u2}
    }
    \subfigure[Heater temperature state (red dashed lines are constraint bounds).]
    {
        \includegraphics[width=0.45\textwidth]{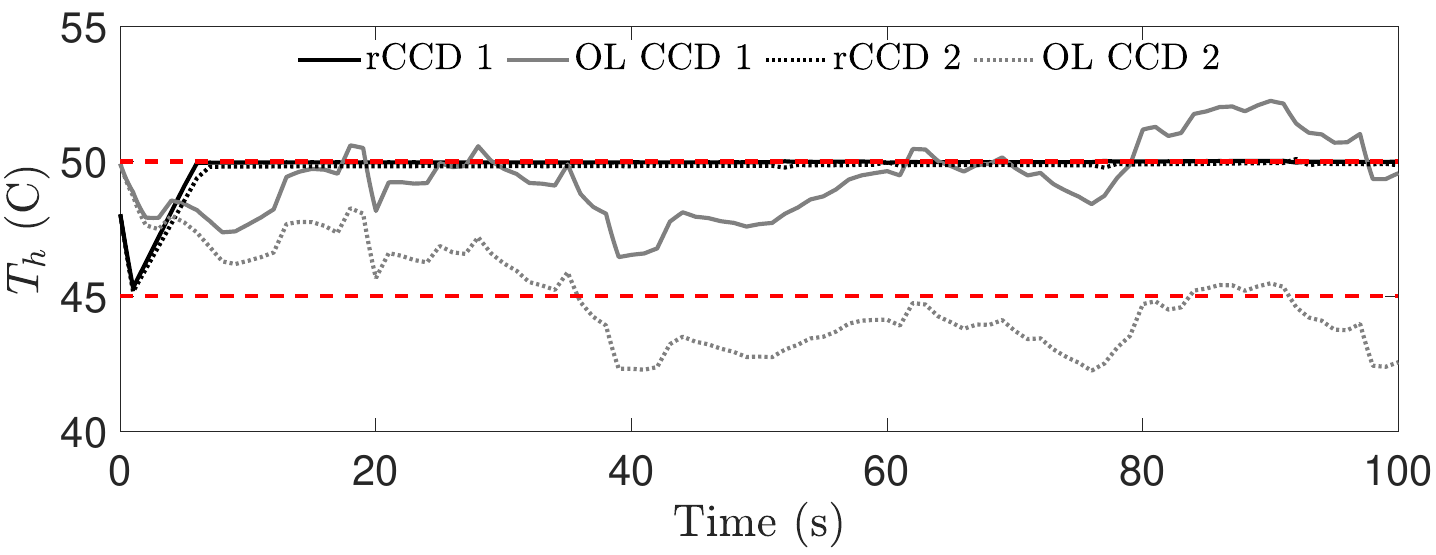}
        \label{fig-cs2_t}
    }
    \caption{\textbf{Case Study 2},  unplanned load profiles:  the pre-optimized rCCD plant/controller is robust to uncertainties in the load profile.}
    \label{fig-cs2}
\end{figure}

Now we consider cases where the profile encountered during system operation is different from the load profile planned for in both the rCCD and OL CCD algorithms. In this case study, the actual loads $\tilde{d}=d+w$ are chosen to be within the uncertainty set $w\in\mathcal{W}$ that Alg.~\ref{alg-rCCD} was designed for. The simulated results are shown in Fig.~\ref{fig-cs2}.

Figure~\ref{fig-cs2_d} gives profiles for $\dot{Q}_h$ and $\dot{m}_e$, Figs.~\ref{fig-cs2_u1}-\ref{fig-cs2_u2} give the optimal control signals, and Fig.~\ref{fig-cs2_t} shows the state temperature $T_h$ for each algorithm. In Fig.~\ref{fig-cs2_d}, the dashed curves represent the load profiles used to optimize the plant/controller designs in the CCD algorithms. The solid curves and dotted curves, conversely, represent the actual load profiles simulated on each optimal plant. Profile $1$ represents a case where the disturbances are higher in magnitude than what was planned for and Profile $2$ represents a case where the disturbances are lower in magnitude than what was planned for. We note that the optimal control signals for the OL CCD case in Figs.~\ref{fig-cs2_u1}-\ref{fig-cs2_u2} are the same for both load profiles since they were completely optimized pre-system build. The solid curves in Fig.~\ref{fig-cs2_t} represent each system's response to Profile 1 and the dotted curves represent the system responses to Profile 2.

The results in Fig.~\ref{fig-cs2} indicate that while a conventional CCD method such as the OL CCD algorithm in Sec.~\ref{sec-OLCCD} cannot meet dynamic constraints in the presence of load uncertainty, the optimal plant/controller design from our proposed rCCD algorithm is robust to load uncertainties. Although both the feedforward control variables and feedback control gains for the system optimized using rCCD were {designed} completely offline, the resulting optimal system is capable of leveraging the corrective feedback input to satisfactorily reject the actual heat load {and drain rate} disturbances encountered while satisfying system state constraints. In other words, we are capable of designing a robust controller that does not require any complex online (real-time) optimization. Moreover, the proposed rCCD algorithm moves beyond the current state-of-the-art in the CCD literature by specifically leveraging robust MPC inside the co-design algorithm itself to account for bounded additive uncertainties.

\FloatBarrier
\section{Conclusions} \label{sec-Conclusions}

In this paper we proposed a nested control co-design algorithm that uses robust model predictive control (rMPC) to reject disturbance uncertainties in closed-loop. 
Plant parameters are optimized in the outer loop of the proposed robust control co-design (rCCD) algorithm. The inner loop utilizes an rMPC approach to account for bounded disturbance uncertainties and design feedforward and feedback gains, thus enabling control variables optimized \emph{offline} by the rCCD algorithm to be used during online operation, even when the disturbance profile differs from the expected one.

In a simulated case study, we applied the proposed approach to a dual-tank thermal management system, which is notionally representative of fuel thermal management systems in aircraft. The proposed algorithm achieved the plant design objective to within 1\% of a benchmark open-loop approach. However, unlike the open-loop approach, the proposed algorithm leveraged the feedback and feedforward gains optimized by the rCCD algorithm to successfully reject an unexpected disturbance signal while satisfying the control objective of maintaining a cold plate surface temperature within defined constraints. Future work will consider formalizing stability and robustness guarantees under recursive linearization of the plant, as well as the use of gain scheduling (rather than time-based scheduling) of the {control variables} to accommodate other plant nonlinearities {and address more disparate disturbance profiles}.


\FloatBarrier
\bibliographystyle{IEEEtran}  
\balance
\bibliography{NSF_Unsolicited_2020.bib}

\end{document}